\documentclass[aps,twocolumn,amsmath,amssymb]{revtex4}
\usepackage{epsf}
\begin{document}

  
\title { The differential sum rule for the 
relaxation rate in the cuprates }  
\author{Ar. Abanov$^1$ and Andrey V. Chubukov$^2$}   
\affiliation{$^1$Los Alamos National Laboratory, MS262, Los Alamos, NM 87545} 
\affiliation{$^2$Department of Physics, University of Wisconsin, Madison, WI 53706}   
\date{\today}   
\begin{abstract}   
Motivated by recent experiments by Basov {\it et al}, 
we study the differential sum rule for the
 effective scattering rate $1/\tau (\omega)$.
We show that in a dirty BCS superconductor,
 the area under $1/\tau (\omega)$ does not change 
between the normal and the superconducting states.
 For magnetically mediated pairing, 
similar result holds between $T<T_c$ and $T \geq T_c$, while
 in the pseudogap phase, 
 $1/\tau (\omega)$ is just suppressed compared to $1/\tau(\omega)$ in the 
 normal state. We 
argue that this violation of the differential sum rule is
 due to the fact that low-energy fermions in the pseudogap phase are still
 incoherent.
 \end{abstract}  
\pacs{PACS numbers:71.10.Ca,74.20.Fg,74.25.-q}   
\maketitle

The analysis of the optical sum rules in condensed matter systems is a 
 valuable tool 
 that helps one to understand the key physics and relevant energy scales
in the problem~\cite{tom}. 
The focus of this report are recent experimental results~\cite{basov}
 for the effective relaxation rate $\tau^{-1} (\omega) =(4\pi/\omega^2_{pl}) Re[1/\sigma (\omega)]$ where 
 $\sigma (\omega) = \sigma_1 (\omega) 
+ i \sigma_2 (\omega)$ is the  optical conductivity, 
$\omega^{2}_{pl} = 4 \pi n e^2/m$ is the plasma frequency, 
and $n$ is the density of particles. 
The data analysis for optimally doped 
$YBa_2Cu_3O_{6.95}$~\cite{basov_1} 
 and $Tl_2Ba_2CuO_{6+x}$~\cite{puchkov} and underdoped 
$YBa_2Cu_4O_8$ and 
$Bi_2Sr_2CaCu_2O_{8+x}$~\cite{basov} 
revealed an approximate differential sum rule
 for $\tau^{-1} (\omega)$ between  $T \geq T_c$  and $T < T_c$: although 
$\int d \omega \tau^{-1} (\omega)$ does not converge, 
it changes very little when the system enters into the superconducting state. 
This differential sum rule, however, is 
{\it not} satisfied between the normal and the 
 pseudogap phases -- $1/\tau (\omega)$ in the pseudogap phase 
appears to be just suppressed.

Quite generally, exact sum rules are related to conservation laws.  
The $f-$sum rule for  the optical conductivity  states 
 that at a given density of particles, 
the total absorbing power of the solid 
characterized by $\sigma_1$ does not depend on the details of the interactions  and is determined only by 
the total number of particles in the system ~\cite{pi_noz}.
The total absorption power is given by $\int_{0}^{\infty} \sigma_1 (\omega)$
Applying the  Kubo formula that
 relates $\sigma (\omega)$ with the  full retarded current-current correlator  $\Pi (\omega)$: 
$\sigma (\omega) = (\omega^2_{pl}/4\pi)~\Pi (\omega)/(-i\omega +  \delta)$, 
separating the frequency integral into the integral over
 infinitesimally small $\omega$ and the rest and using 
 the Kramers-Kronig relation for  $\Pi (\omega) -1$ that 
 {\it vanishes} at the highest frequencies, we  obtain
$\int_{0}^{\infty} \sigma_1 (\omega) =\omega^2_{pl}/8$, independent of the 
 form of $\Pi (\omega)$. 

Is there an analogous sum rule for $1/\tau (\omega)$?
 Using $1/\tau (\omega) = - Im [\omega^2/\Pi (\omega)]/\omega$ and 
 applying the Kramers-Kronig relation, we find
\begin{equation}
\int_0^{\infty} \frac{d \omega}{\tau (\omega)} = 
\frac{\pi}{2} \left[\Re \frac{\omega^2}{\Pi (\omega)}_{\omega \rightarrow 0} +
 C\right] = \frac{\pi}{2}~C
\label{3}
\end{equation}
The constant $C$ again has to be chosen such that 
$\omega^2/\Pi (\omega) + C$ vanishes at $\omega \rightarrow \infty$.
 However, $C$  turns out to be infinite 
as at high frequencies $\Pi (\omega) \approx 1$, and 
$\omega^2/\Pi (\omega)$ diverges. 
This divergence implies  that 
there is no 
conservation law associated with 
 the relaxation rate and hence 
{\it no sum rule for $1/\tau (\omega)$}. 

Still, one can try to see whether one can get useful information by 
comparing $1/\tau (\omega)$ for two different system parameters, e.g. 
temperatures, which do not affect the system behavior at high frequencies. 
Suppose momentarily that $1/\tau (\omega, T_1) - 1/\tau (\omega, T_2)$ vanishes at high frequencies. The Kramers-Kronig relation 
 is then  fully satisfied and using (\ref{3}) 
we immediately find that $\int d \omega (\tau^{-1} (\omega, T_1) - \tau^{-1} (\omega, T_2)) =0$, i.e., 
the area under $1/\tau (\omega)$ does not change with $T$. 
The conservation of the total area under $1/\tau (\omega)$ 
would create a valuable tool to study 
 the evolution of the spectral weight  between, e.g., 
 the normal and the superconducting states. 
This new differential sum rule, however, is  not 
associated with a conservation law and therefore is not guaranteed to 
be  satisfied --
 only explicit calculations  can determine whether or not
the temperature dependence in $\Pi (\omega, T)$ is weak enough to ensure the 
 convergence of   $1/\tau (\omega, T_1) - 1/\tau (\omega, T_2)$.
\begin{figure}[tbp]
\begin{center}
\epsfxsize=\columnwidth 
\epsffile{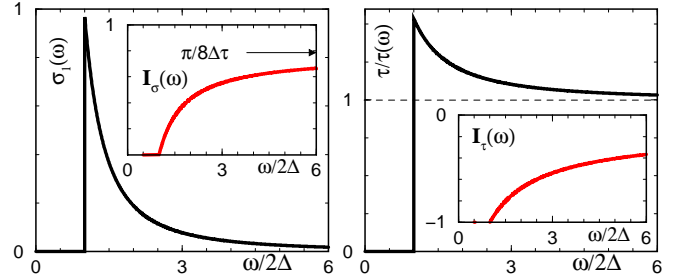}
\end{center}
\caption{The results for $\sigma_1 (\omega)$ 
and $1/\tau(\omega)$ (in arbitrary units) 
 for a  BCS superconductor, to first order in $1/\tau \Delta$ (clean limit).
In the dirty limit, the jump in the conductivity at $\omega = 2\Delta$ is 
much smaller. 
The inserts show 
$I_\sigma (\omega) = 8/\omega^2_{pl} \int_0^\omega dx \sigma_1 (x)$ and 
$I_\tau (\omega) = (\tau/2\Delta) \int_{0}^\omega d x (1/\tau(x) -1/\tau)$.
 The arrow indicates the  value of $I_\sigma (\infty)$ and . 
The differential sum rule for $1/\tau (\omega)$ is satisfied if 
 $I_\tau (\infty) =0$.}
\label{fig2}
\end{figure}
In this communication we study under which conditions
 the differential sum rule for $1/\tau (\omega)$  actually holds,
  and  at which frequencies it is exhausted. We consider the magnetic scenario
 for the pairing in the cuprates, and argue  that although 
 the differential sum rule is approximately satisfied for all $T$, 
 it is exhausted at energies comparable to the pairing gap only
 if there is a strong  
feedback effect from the pairing on the fermionic propagator -- otherwise,
 the sum rule is exhausted  at much larger frequencies. We associate the
 first regime with $T < T_c$, and the second one with the pseudogap phase.
    
To put our analysis of the spin mediated pairing into perspective we 
 first  analyze the situation 
 in a dirty BSC  $s-$wave superconductor at $T=0$, when the pairing causes a
 strong feedback on the fermionic propagator, and in the 
 toy model where there is no feedback from the pairing on the fermionic self-energy.

 The theory of a dirty superconductor is well developed~\cite{ag,anderson}.
 In the normal state, the inelastic scattering by impurities yields a retarded 
fermionic self-energy $\Sigma (\omega) = i/2\tau $, In a superconducting state,
 this self-energy is modified due to a feedback from superconductivity and takes the form $\Sigma (\omega) = (i/2\tau) \omega/\sqrt{\omega^2 - \Delta^2}$ where $\Delta$ is the superconducting gap~\cite{ag}. 
Substituting these forms into the current-current polarization bubble and performing the momentum integration 
 we obtain
\begin{eqnarray}
&&\Pi (\omega) =\int_0^{\infty} d \Omega~\frac{1}{\left(\sqrt{\Omega^2_+ - \Delta^2} +  \sqrt{\Omega^2_- - 
\Delta^2} + i/\tau\right)} 
 \nonumber\\
&&\times 
\frac{\sqrt{\Omega^2_+ - \Delta^2} \sqrt{\Omega^2_- - \Delta^2} - \Delta^2 -
 \Omega_+ \Omega_-}{\sqrt{\Omega^2_+ - \Delta^2} \sqrt{\Omega^2_- - \Delta^2} 
}
\label{4}
\end{eqnarray}
where $\Omega_{\pm} = \Omega \pm \omega/2$. In the normal state, this reduces to a conventional Drude form $\Pi (\omega) = \omega/(\omega +  i/\tau)$.
In the superconducting state, 
 the frequency integral in (\ref{4}) can be 
 evaluated analytically in the clean limit $\Delta \tau \gg 1$.
 After lengthy but 
 straightforward calculations we found that both $\sigma_1 (\omega)$ and 
$1/\tau (\omega)$ vanish at $\omega <2 \Delta$, while at larger frequencies 
\begin{equation}
\sigma_1 (\omega) = \frac{\omega^2_{pl}}{4\pi\tau \omega^2}
 E\left(\sqrt{1 - \frac{4\Delta^2}{\omega^2}}\right), \,
\frac{1}{\tau(\omega)} = 
\frac{4\pi\sigma_1 (\omega) \omega^2}{\omega^2_{pl}} 
\label{5}
\end{equation}
where $E(x)$ is the complete elliptic integral 
of the second kind~\cite{morr}. 
At $\omega = 2\Delta +0$, $E = \pi/2$ and both 
$\sigma_1 (\omega)$ and $1/\tau (\omega)$ jump to finite values.
At high frequencies, $E (x \approx 1) \rightarrow 1$, 
$\sigma_1 (\omega)$ vanishes as $\omega^{-2}$, and 
$1/\tau (\omega)$ approaches the normal state result $\tau(\omega) = \tau$.
 To the same order, we also have $\Pi (0) = 1 - \pi/(8\Delta \tau)$. 
 We checked analytically that the $f-$sum rule 
$(8/\omega^2_{pl}) \int_{+0}^{\infty} d \omega \sigma_1 (\omega)= 
1 -\Pi (0)$ is indeed satisfied.

Expanding  $E(x)$ near $x=1$, we find that at high frequencies 
$\tau^{-1} (\omega) -
\tau^{-1} \approx (2\Delta^2/\omega^2\tau) (\log(2\omega /\Delta) -0.5)$, i.e.
 $\int d \omega (1/\tau (\omega) - 1/\tau)$ converges. The convergence implies
 that for a dirty BCS superconductor, 
the differential sum rule for $1/\tau (\omega)$ is an exact one.

The plots of $\sigma_1 (\omega)$ and 
$1/\tau (\omega)$  are presented in Fig.~\ref{fig2} together with the results 
 for $I_\sigma (\omega) = (8/\omega^2_{pl})
 \int_0^\omega dx \sigma_1 (x)$ and 
$I_\tau (\omega) = (\tau/2 \Delta) \int_{0}^\omega 
d x (1/\tau(x) -1/\tau)$. It follows from (\ref{5}) that   
 the sum rules for $\sigma_1 (\omega)$ and $1/\tau (\omega)$ should be
 exhausted  at frequencies of the order of $\Delta$. 
  Fig.~\ref{fig2} shows that this is indeed the case.
\begin{figure}[tbp]
\begin{center}
\epsfxsize=\columnwidth 
\epsffile{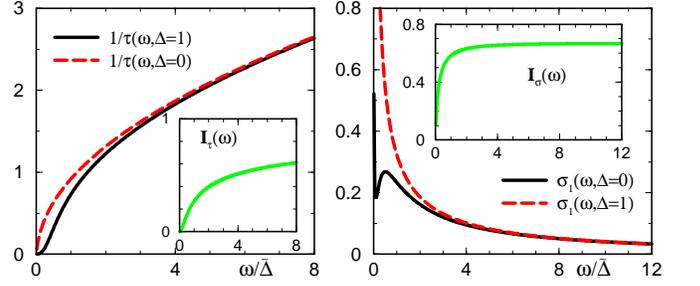}
\end{center}
\caption{The results for  $1/\tau (\omega)$ (a) and 
$\sigma_1 (\omega)$ (b) 
 for a toy model in which the pairing is not accompanied by 
 the feedback on the electrons. The frequency is in the units of 
${\tilde \Delta}$ (the peak frequency 
in the DOS).  We used ${\tilde \Delta} = 0.25 {\bar \omega}$.
 Observe that $1/\tau (\omega)$ is nearly homogeneously 
 suppressed at $\omega = O({\tilde \Delta})$, and the 
differential $I_\tau (\omega)$ converges to zero only at 
 $\omega \sim 10^3 {\tilde \Delta}$ (not shown). 
On the other hand,  $I_\sigma (\omega)$
 converges to $I_\sigma (\infty) \approx 0.67$ 
already at $\omega \sim 4 \Delta$.}
\label{fig4}
\end{figure}
We next consider the behavior of 
 $\sigma_1 (\omega)$ and $1/\tau (\omega)$ in the    
 toy model  in which we assume 
 that the pairing does not change the fermionic self-energy.
  Having the cuprates in mind, we also assume, 
 that the normal state is not a Fermi liquid, i.e., fermionic
 self-energy at low frequencies behaves as $\Sigma (\omega) = 
(i \omega)^{\alpha} {\bar \omega}^{1-\alpha}$ with $\alpha <1$, 
 
For the frequency dependent self-energy $\Sigma (\omega)$,  
the current-current correlator $\Pi (\omega)$ is still 
given by Eq. (\ref{4}), but now $\Omega$ is substituted by 
$\Omega + \Sigma (\Omega)$. 
 For definiteness, we present the results for $\alpha = 1/2$, 
which is the normal state 
quantum-critical exponent  in the spin-fermion theory~\cite{abanov01}, but
 the results are qualitatively the same for all $\alpha$ including the
 marginal Fermi liquid limit $\alpha \rightarrow 0$~\cite{av}. 
For $\alpha =1/2$, the fermionic density of states $N(\omega) = 
Im [{\tilde \Sigma} (\omega)/(\Delta^2 - {\tilde \Sigma}^2 (\omega))^{1/2}]$ 
has a maximum at $\omega = {\tilde \Delta} \approx \Delta^2/{\bar \omega}$, 
 that could  be identified with the pseudogap. Still, 
 $N(\omega)$ remains finite at all frequencies.   
The assumption that the self-energy does not change is internally consistent if indeed  the 
anomalous normal state behavior extends to frequencies which well  exceed the pairing gap, i.e., when ${\bar \omega} \gg {\tilde\Delta}$.
 
Evaluating $\Pi (\omega)$ and substituting it into  
$\sigma_1 (\omega)$ and $1/\tau (\omega)$ we found that 
in  the normal state,  $\sigma_{1,n} (\omega) \propto 
(\omega {\bar \omega})^{-1/2}$ at 
$\omega \ll {\bar \omega}$ and $\sigma_{1,n} \propto 
({\bar \omega}/\omega^3)^{1/2}$ at
 $\omega \gg {\bar \omega}$, while $1/\tau_n (\omega) 
\propto (\omega {\bar \omega})^{1/2}$ in both limits. 
For $\Delta \neq 0$, we found that for $\omega \ll {\tilde \Delta}$,
$\sigma_1 (\omega) \propto (\omega {\bar \omega})^{-1/2} 
(\omega/{\tilde \Delta})^{5/2}$ and $1/\tau (\omega) \propto 
(\omega {\bar \omega})^{1/2}  (\omega/{\tilde \Delta})^{7/2}$.
We see that both $\sigma_1 (\omega)$ and $1/\tau (\omega)$ 
are reduced compared to their normal state values but are still finite.
At $\omega \sim {\tilde \Delta}$, both $\sigma_1 (\omega)$ and 
$1/\tau (\omega)$ become comparable to the normal state results. 
At larger $ {\tilde \Delta} \ll \omega \ll {\bar \omega}$, 
$\sigma_1 (\omega) = \sigma_{1,n} (\omega) - 
1.992 (\omega^2_{pl}/4\pi) {\tilde \Delta}(\omega^3 {\bar \omega})^{-1/2}$ 
and $1/\tau (\omega) = 1/\tau_n (\omega) - 
3.51 {\tilde \Delta} ({\bar \omega}/\omega)^{1/2}$.
Finally, at $ \omega \gg
 {\bar \omega}$,  $\sigma_1 (\omega) - \sigma_{1,n} (\omega) 
\propto \omega^{-7/2}\log{\omega}$ and 
$1/\tau (\omega) - 1/\tau_n (\omega) \propto  \omega^{-3/2} \log{\omega}$.  

\begin{figure}[tbp]
\begin{center}
\epsfxsize=\columnwidth 
\epsffile{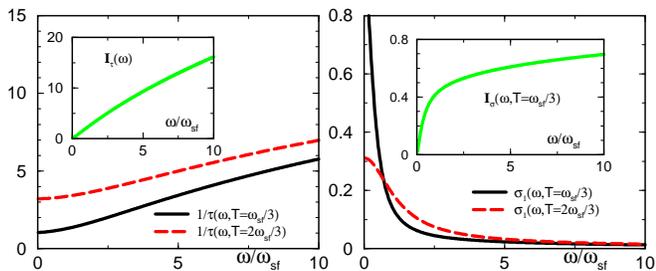}
\end{center}
\caption{The normal state 
results for  $1/\tau (\omega)$ (a) and 
$\sigma_1 (\omega)$ (b)  
for the spin-fermion model for $\lambda =2$. 
The insets show $I_\sigma (\omega)$ and the differential $I_\tau (\omega)$ 
 between $T = \omega_{sf}/3$ and
 $T = 2 \omega_{sf}/3$.  The sum rule for $1/\tau (\omega)$ is not satisfied due to weak convergence at high frequencies.   
 $I_\sigma (\omega)$ flattens at  $\omega \sim 10 {\omega_{sf}}$ 
 but converges to the $f-$sum rule value $I_\sigma (\infty) =1$
 only at extremely  high 
$\omega \sim 10^3 \omega_{sf}$ (not shown). 
} 
\label{fig3}
\end{figure}
We see that $\sigma_1 (\omega)$ converges to its normal state value at frequencies of order ${\tilde \Delta}$, as in a dirty BCS superconductor;
 the  sum rule for $\sigma_1 (\omega)$ is then  exhausted  at $\omega \sim  {\tilde \Delta}$. 
This behavior is illustrated in Fig.~\ref{fig4}a where we present 
the results of our numerical calculations 
-- $I_\sigma (\omega)$ converges to  $I_\sigma (\infty) = 1 - \Pi (0)$ ($\approx 0.67$ for our choice of 
${\bar \omega} = 2 \Delta$) already at $\Omega \sim {\tilde \Delta}$.
On the other hand,  
 $\tau^{-1} (\omega) - \tau_n^{-1} (\omega)$ scales as $ \omega^{-1/2}$ between $\omega \sim {\tilde \Delta}$ and $\omega \sim {\bar \omega}$ 
 such that at these frequencies 
 $I_\tau (\omega) = \int d \omega (\tau^{-1} (\omega) - \tau_n^{-1} (\omega))$  does not converge. 
Furthermore, at these frequencies, $1/\tau (\omega)$ is still 
 {\it smaller} than $1/\tau_n (\omega)$.
 This result holds for all $\alpha <1$ as one can straightforwardly verify.
 Only at $\omega > {\bar \omega}$,   $\tau^{-1} (\omega)$ finally 
 becomes larger than
 $\tau_n^{-1} (\omega)$, and 
$I_{\tau} (\omega)$  converges. 
The convergence  implies 
that the differential sum rule for 
$1/\tau (\omega)$ is again exactly satisfied, however {\it 
it is exhausted only at frequencies that well exceed the pseudogap}.
We present the numerical 
 results for  $1/\tau (\omega)$ and $I_\tau (\omega)$ in  Fig.~\ref{fig4}b.

We now present the results for  $\sigma_1 (\omega)$ and 
$1/\tau (\omega)$ for spin-fluctuation mediated $d-$wave  pairing. 
We obtained these results by 
 solving a set of coupled Eliashberg equations 
for the  
spin-fermion model that describes the spin 
fluctuation exchange at low energies \cite{acs}.
We will demonstrate that at low $T$, the behavior of the conductivity and the relaxation rate resembles that in a dirty BSC superconductor, while 
 immediately below the pairing instability the system behavior is similar to that in the toy model for the pseudogap.
 
The spin-fermion model is 
 characterized by a single dimensionless coupling constant 
$\lambda$ and a single overall energy ${\bar \omega}$ 
 that scales with the effective spin-fermion interaction~\cite{abanov01}.
We will also use a characteristic energy scale for the spin fluctuations
$\omega_{sf} = \bar{\omega }/4\lambda ^{2}$.
 A fit to the NMR, ARPES and  neutron experiments 
yields $\lambda \sim 1-2$ near the optimal doping~\cite{abanov01}. 
 We refer the readers to Ref~\cite{abanov01} for the
discussion of the applicability of the model to the cuprates 
 and  the justification of the  Eliashberg approach at
 strong spin-fermion coupling despite the formal absence of the 
 Migdal theorem. 

We begin with the normal state.
In Fig.~\ref{fig3}a we present our results for $1/\tau (\omega)$ and 
$I_{\tau }(\omega )$ at various $T$. 
 For definiteness we used $\lambda =2$.
We see that 
$I_\tau (\omega)$ diverges at high frequencies, i.e., the differential sum rule is {\it not} satisfied. We
checked analytically that this is caused by the $1/\omega$ behavior of the
 integrand in $I_\tau (\omega)$.
%

For completeness, in 
Fig.~\ref{fig3}b we present the results for $\sigma_1 (\omega)$ 
and $I_\sigma (\omega)$ at various $T$.
 We see that $I_\sigma (\omega)$  
 flattens up at $\omega \geq 10 \omega_{sf}$, 
but its value is still by about 
  $30\%$ smaller than it should be
 for $\omega = \infty$.  The full sum rule is exhausted
 only at unrealistically large 
$\omega \sim 10^3 \omega_{sf}$ (Ref.~\cite{abanov01}) 
where the low-energy theory is clearly inapplicable.
The weak convergence of $I_\sigma (\omega)$ is 
 related to the fact that over a wide frequency 
range $\sigma_1 (\omega)$  
is inversely proportional to $\omega$, and 
$I_\sigma (\omega)$ increases as  $\log \omega$.
The practical application of this result is that the plasma frequency $\omega_{pl}$ obtained by integrating the measured $\sigma_1 (\omega)$ 
up to, say, $1eV$  should be about  $30\%$
less than the actual plasma frequency.
This is consistent  with the fact that to match the measured
 $\sigma_1 (\omega \rightarrow 0)$ with the theoretical 
 $\sigma_1 (\omega \rightarrow 0)$ 
 obtained  using $\Sigma (\omega)$ extracted from the 
ARPES data, one needs the plasma frequency of about $20000 cm^{-1}$~\cite{abanov01,millis}, 
 while $\omega_{pl}$ extracted from the
 ``partial'' sum rule~\cite{puchkov} is around $16000 cm^{-1}$.

We next consider what happens below the pairing 
instability temperature $T_{ins} \sim 0.2  {\bar \omega}$~\cite{acf}.
Earlier  Schmalian and the two of us have found~ \cite{acs}  that 
 at $T \leq T_{ins}$, the  
the fermionic self-energy  remains large at the smallest $\omega$ 
  and gradually evolves from its normal state value.
 This gradual behavior is qualitatively different from a dirty BCS 
superconductor, as in the latter the quasiparticle spectral function 
 and the density of states instantly drop to zero  at frequencies 
below $\Delta$ due to a feedback from the pairing~\cite{ag,anderson}.   
We found in ~\cite{acs} that the behavior similar to that in a BSC superconductor occurs only below $T_c < T_{ins}$, and the difference between $T_{ins}$ 
 and $T_c$ increases with increasing $\lambda$. We conjectured that 
 at $T_c < T < T_{ins}$, fluctuations destroy coherent superconductivity, i.e.,  the system is in the pseudogap regime.

\begin{figure}[tbp]
\begin{center}
\epsfxsize=\columnwidth 
\epsffile{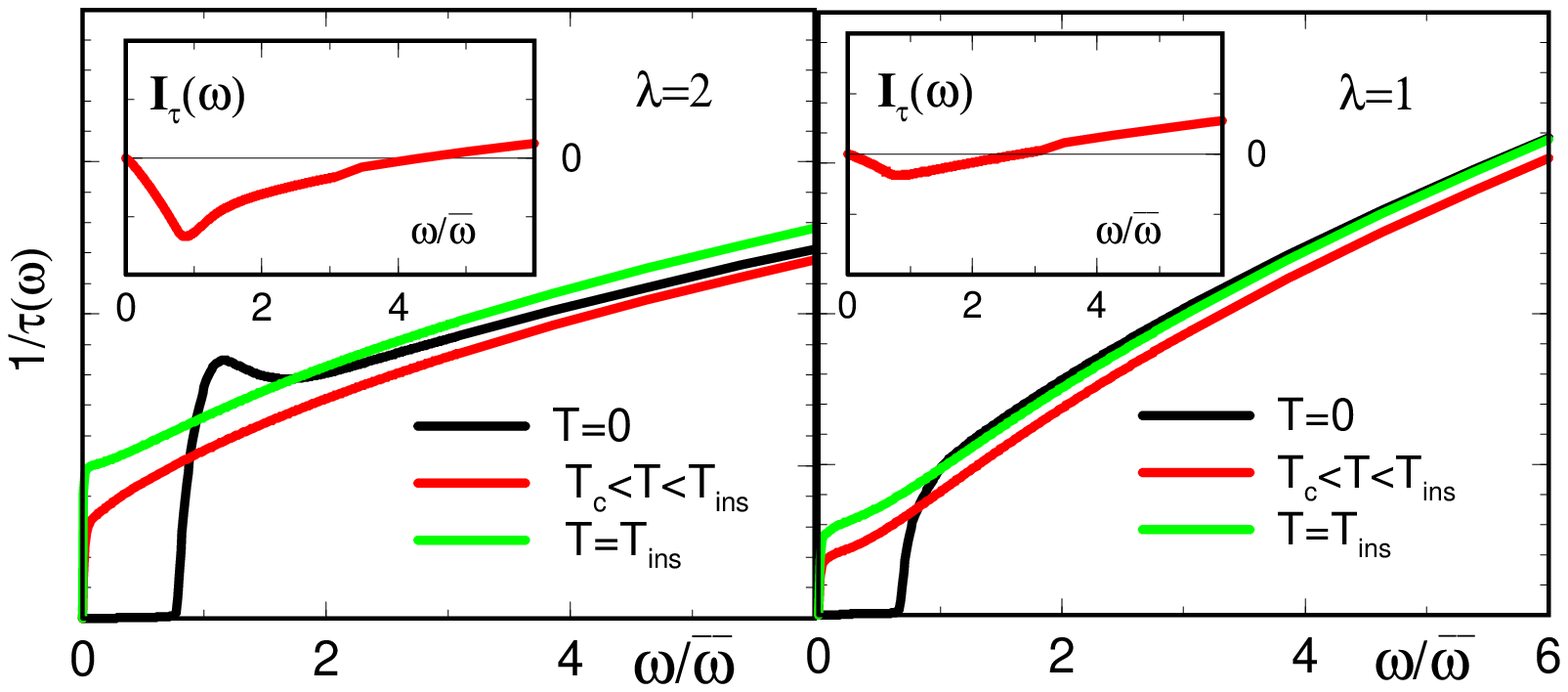}
\epsfxsize=\columnwidth 
\epsffile{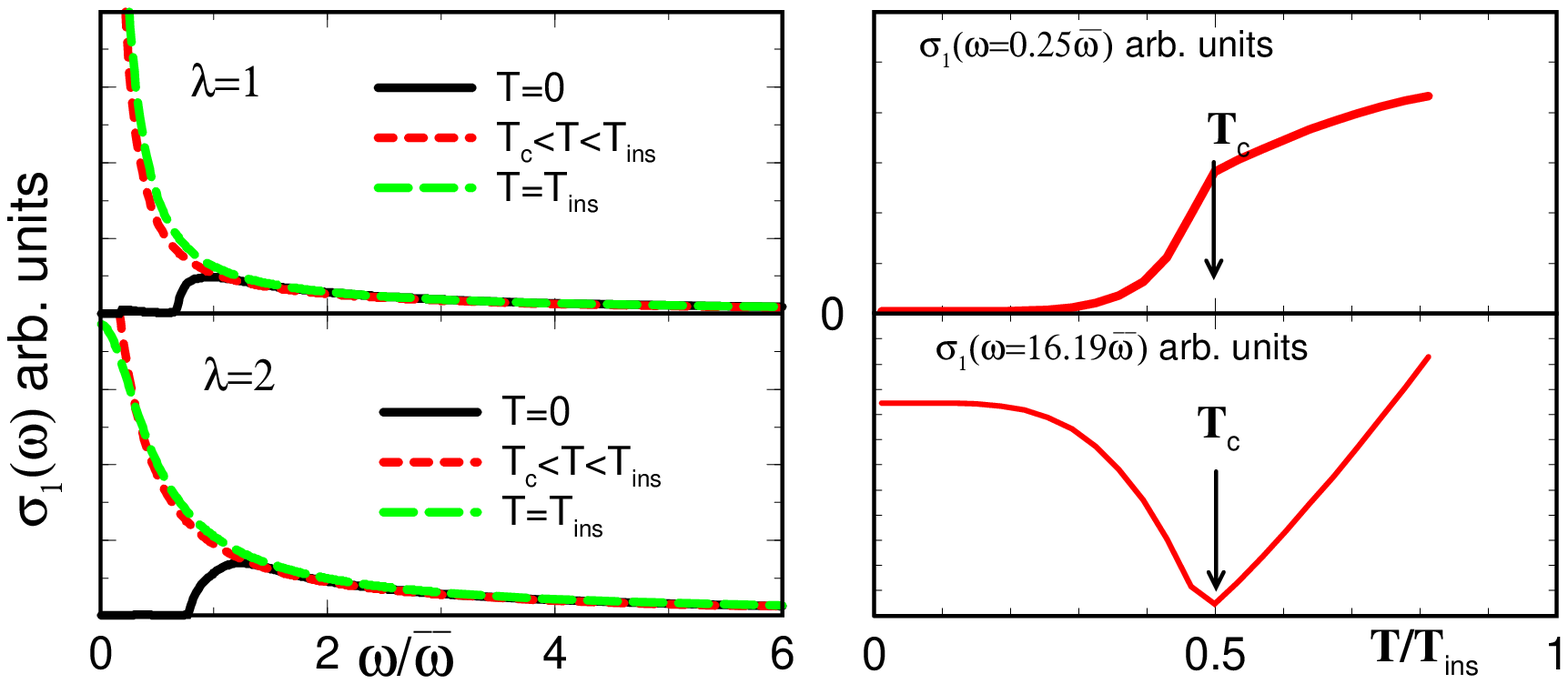}
\end{center}
\caption{ 
The behavior of $1/\tau(\omega )$ and 
 $\sigma_{1}(\omega)$ 
in the spin-fermion model  below
 the pseudogap temperature $T_{ins}$
 for $\lambda =2$ ($\Delta \sim 0.3 {\bar \omega}$) and 
$\lambda =1$ ($\Delta \sim 0.2 {\bar \omega}$). 
Observe that the overshoot between the spectra of $1/\tau (\omega)$ 
 develops only below $T_c$. 
 The inserts in the upper panels show the behavior of 
the differential sum rule $I_{\tau }(\omega )$  
between $T< T_c$ and $T_c< T <T_{ins}$.  
 The differential sum rule is not exact, but is approximately 
satisfied if $\omega \sim  3-4{\bar \omega} \sim 10 \Delta$.
The right lower panels  show the behavior 
of $\sigma_1 (\omega)$ vs $T$  at small 
 and large frequencies for $\lambda =1$. Observe that the 
  changes in $\sigma_1$ are confined to $T_c$ rather than to $T_{ins}$. 
}
\label{fig33}
\end{figure}

In Fig.~\ref{fig33} we present the results for $\sigma_1 (\omega)$ and 
$1/\tau (\omega)$ for two different $\lambda $ and 
for three different temperatures: 
$T \ll T_c$, 
 $T_c < T < T_{ins}$ and $T=T_{ins}$ where $\sigma_1 (\omega)$ and 
$1/\tau (\omega)$ are the same as in the normal state.
We clearly see that between   $T_c < T < T_{ins}$, 
$1/\tau (\omega)$  is nearly homogeneously suppressed, while 
 at smaller  $T$  
 it shifts  from the lowest frequencies to $\omega \sim 2\Delta$ 
($\sim 0.8 {\bar \omega}$) and develops an overshoot at $\omega \geq 2\Delta$.
  The magnitude of the overshoot  depends on the coupling and is larger
 at larger $\lambda$.
We compared our results with 
 the fermionic self-energy~\cite{acs}, 
and found that the temperature
 where $1/\tau (\omega)$ develops an overshoot, agrees with the onset 
temperature for the feedback effects on the fermionic propagator.

For quantitative analysis of the sum rule  
 we looked analytically into the high frequency behavior of 
  $1/\tau (\omega)$. 
  We found that the temperature variation of $1/\tau (\omega)$ 
at high frequencies is still strong enough to prevent  
the  differential sum rule for $1/\tau$. 
On the other hand, we see from the insert in Fig.~\ref{fig33}b that 
 this variation  can be safely neglected  if the frequency 
integration is restricted to $\omega \sim 3-4{\bar \omega} 
\sim 10 \Delta$  
(for optimally doped $YBCO$, $10 \Delta \sim 
2500 cm^{-1}$).
In Fig. \ref{fig33}b we show the results for the differential sum 
rule between $T \ll T_c$ and $T_c < T < T_{ins}$. 
 We see that  the spectral weight is almost exactly conserved if 
the upper limit of the frequency integral is chosen close to 
$3-4 {\bar \omega}$ . 
 The conservation of the spectral weight between $T \ll T_c$ and 
$T \geq T_c$, and the apparent loss  of the spectral weight between 
$T_c$ and $T_{ins}$ are
the main results of recent experimental analysis~\cite{basov}.
 We see that these results are fully reproduced in our analysis.

We also see from Fig.~\ref{fig33} that
 at small $\omega$,  $\sigma_1 (\omega)$ keeps increasing 
between $T_{ins}$ and $T_c$. This indicates that 
the development of the pseudogap  does not give rise to a suppression of 
the conductivity at the lowest frequencies. The latter is
 only reduced  below $T_c$. 
The sensitivity of $\sigma_1 (\omega =0)$ to $T_c$ rather than to
the pseudogap temperature is  consistent with the data~\cite{nicole}. 
 The low frequency behavior of $\sigma_1 (\omega=0)$ below $T_c$ is not captured in our theory as it is predominantly
 determined by impurities~\cite{lee}.   
Finally, we found both analytically and numerically that 
  at large $\omega$, $\sigma_1 (\omega)$ also 
turns out to be sensitive to $T_c$  (see the inset in Fig.~\ref{fig33}a).
 This also agrees with the data~\cite{vdm}. 

To conclude, in this paper we considered the differential  sum rule
 for the effective scattering rate $1/\tau (\omega)$ (the difference between the area under $1/\tau (\omega)$ for two different temperatures).
 We argued that for spin-fluctuation mediated pairing, 
this sum rule is generally not an exact one, but is rather 
well satisfied below $T_c$ and  is exhausted  at frequencies compared to
 the pairing gap, $\Delta$. We identified this behavior with the strong feedback from the pairing on the fermionic self-energy. We found that in the 
 pseudogap region, where feedback effects are small, the differential 
sum rule is (approximately) 
 exhausted only at much larger energies, comparable to the upper 
limit of non-Fermi-liquid behavior in the normal state, while at 
$\omega = O(\Delta)$, $1/\tau (\omega)$ is nearly homogeneously 
suppressed compared to the normal state.
We argued that this behavior as well as the behavior of 
$\sigma_1 (\omega)$ are consistent with the experimental 
data on the cuprates. 

It is our pleasure to thank D. N. Basov, G. Blumberg, N. Bontemps,
 C. Homes, D. van der Marel, A. Millis, M. Norman,  D. Pines, and J. Tu 
 for useful conversations. We are also thankful to D. N. Basov,  N. Bontemps,
 and J. Tu  for sharing
 the unpublished results with us. 
The research was supported by NSF DMR-9979749 (A. Ch.) 
and by DR Project 200153 
at Los Alamos National Laboratory (Ar. A.).


\begin{thebibliography}{99}

\bibitem{tom} T. Timusk and B. Statt, Rep. Prog. Phys. {\bf 62}, 61 (1999). 

\bibitem{basov} D.N. Basov, E.J. Singley and S.V. Dordevic, cond-mat/0103507.

\bibitem{basov_1} D.N. Basov et al. \prl {\bf 77}, 4090 (1996).

\bibitem{puchkov} 
 A.V. Puchkov,  D.N. Basov, and T. Timusk, 
J. Cond. Matt. Phys. {\bf 8}, 10049 (1996).

\bibitem{pi_noz} D. Pines and Ph. Nozi\`eres, 
{\it The Theory of Quantum Liquids}, 
Cambridge, Massachusetts, Perseus books (1999).

\bibitem{ag} A.A. Abrikosov A. A. and L.P. Gor'kov, 
Sov. Phys. JETP, {\bf 8}, 1090 (1959).

\bibitem{anderson}  P.W. Anderson J. Phys. Chem. 
Solids, {\bf 11}, 26 (1959). 

\bibitem{morr} see also 
H. Westfahl and D. Morr, cond-mat/0002039.

\bibitem{abanov01}  
A. Abanov, A. V. Chubukov, J. Schmalian, 
cond-mat/0107421; A. Chubukov, J. Schmalian and D. Pines 
 cond-mat/xxxx, and references therein. 

\bibitem{av}  E. Abrahams and C. Varma, \prl 
{\bf 84}, 4652 (2001) and references therein.

\bibitem{millis} N. Shah and A.J. Millis, cond-mat/0104502.

\bibitem{acf} Ar. Abanov, A. V. Chubukov, and A. M. Finkel'stein, 
Europhys. Lett., {\bf 54} ,488-494 (2001).

\bibitem{acs} A. Abanov, A. V. Chubukov, and J. Schmalian, 
Europhys. Lett.{\bf  55} 369 (2001).

\bibitem{nicole} A.F. Santander-Siro et al, cond-mat/0107161;  
J.J. Tu et al, cond-mat/0107349.

\bibitem{lee} P.A. Lee, \prl {\bf 71}, 1887 (1993);
 S. M. Quinlan, P. J. Hirschfeld, D. J. Scalapino, 
\prb {\bf 53}, 8575 (1996).

\bibitem{vdm} D. van der Marel, Physica C {\bf 341-348}, 1531 (2000). 

\end{thebibliography}
\end{document}